\documentclass[12pt,preprint]{aastex}

\shorttitle{Exoplanets imaging with a Phase-Induced Amplitude Apodization Coronagraph}
\shortauthors{O. Guyon et al.}
\usepackage{graphicx}
\usepackage[mathscr]{eucal}
  
\begin{document}

\title{Exoplanets imaging with a Phase-Induced Amplitude Apodization Coronagraph - I. Principle}
       
\author{Olivier Guyon, Eugene A. Pluzhnik, Raphael Galicher, Frantz Martinache}
\affil{Subaru Telescope, National Astronomical Observatory of Japan}
\affil{650 North A'ohoku Place, Hilo, HI 96720 USA}
\email{guyon@subaru.naoj.org}

\author{Stephen T. Ridgway}
\affil{National Optical Astronomical Observatories}
\email{ridgway@noao.edu}

\author{Robert A. Woodruff}
\affil{Lockheed Martin Space Corporation}
\affil{P.O. Box 179, Denver CO 80201-0179}

\begin{abstract}
Using 2 aspheric mirrors, it is possible to apodize a telescope beam without losing light or angular resolution: the output beam is produced by ``remapping'' the entrance beam to produce the desired light intensity distribution in a new pupil. We present the Phase-Induced Amplitude Apodization Coronagraph (PIAAC) concept, which uses this technique, and we show that it allows efficient direct imaging of extrasolar terrestrial planets with a small-size telescope in space. The suitability of the PIAAC for exoplanet imaging is due to a unique combination of achromaticity, small inner working angle (about 1.5 $\lambda/d$), high throughput, high angular resolution and large field of view. 3D geometrical raytracing is used to investigate the off-axis aberrations of PIAAC configurations, and show that a field of view of more than 100 $\lambda/d$ in radius is available thanks to the correcting optics of the PIAAC. Angular diameter of the star and tip-tilt errors can be compensated for by slightly increasing the size of the occulting mask in the focal plane, with minimal impact on the system performance. Earth-size planets at 10 pc can be detected in less than 30s with a 4m telescope. Wavefront quality requirements are similar to classical techniques.
\end{abstract}


\section{Introduction}
With now more than 100 exoplanets known, the scientific interest for direct detection of exoplanets is very high: unlike indirect detection techniques, it will allow characterization through spectroscopy. Of particular importance is the discovery and characterization of planets similar to ours, which are the prime candidates for detectable evidence of life outside our solar system.

The point spread function (PSF) obtained by a telescope pupil, even in the absence of aberrations, is poorly suited for high contrast imaging, and most high dynamical range imaging concepts therefore propose to feed the telescope beam to a coronagraph \cite{rodd97,roua00}. Another approach to high dynamical range imaging is to ``shape'' the pupil illumination function to produce a high contrast PSF: this technique is referred to as pupil apodization \cite{jacq64}. We will refer to these techniques as classical coronagraphy and classical apodization.

Properly apodized pupils are suitable for high dynamical range imaging. Various apodization functions or pupil shapes have been suggested to produce a PSF with very dark areas at small angular distances \cite{jacq64,kasd03,vand03}. The apodization technique offers some unique advantages over most coronagraphs:
\begin{itemize}
\item{It is simple and robust. It can be easily made achromatic if a binary transmission mask is used.}
\item{It is insensitive to the stellar angular size, and tolerates small telescope pointing errors.}
\end{itemize}

Unfortunately, pupil apodization is usually performed by selectively absorbing light in the pupil plane (classical pupil apodization, or CPA), which reduces both the angular resolution (typically by a factor 3) and throughput (by about a factor 4 to 10) of the telescope. Moreover, only a fraction of the field of view is usable with some apodization masks. These effects are stronger as the required PSF contrast is increased.


Alternatively, pupil apodization may be performed by geometrical remapping of the flux in the pupil plane. This technique has been commonly used for shaping the beam of lasers and to reduce the side lobes of an antenna in radio astronomy. Optical imaging with a remapped pupil was first suggested for sparse interferometric arrays to produce a single diffraction peak PSF \cite{labe96,bocc00,guyo02,riau02}. Guyon (2003) showed that a similar method (Phase-Induced Amplitude Apodization, PIAA) can be applied on a single aperture telescope for high dynamical range imaging. PIAA combines the advantages of pupil apodization listed above with full throughput and no loss of angular resolution.

\begin{figure}[htb]
\plotone{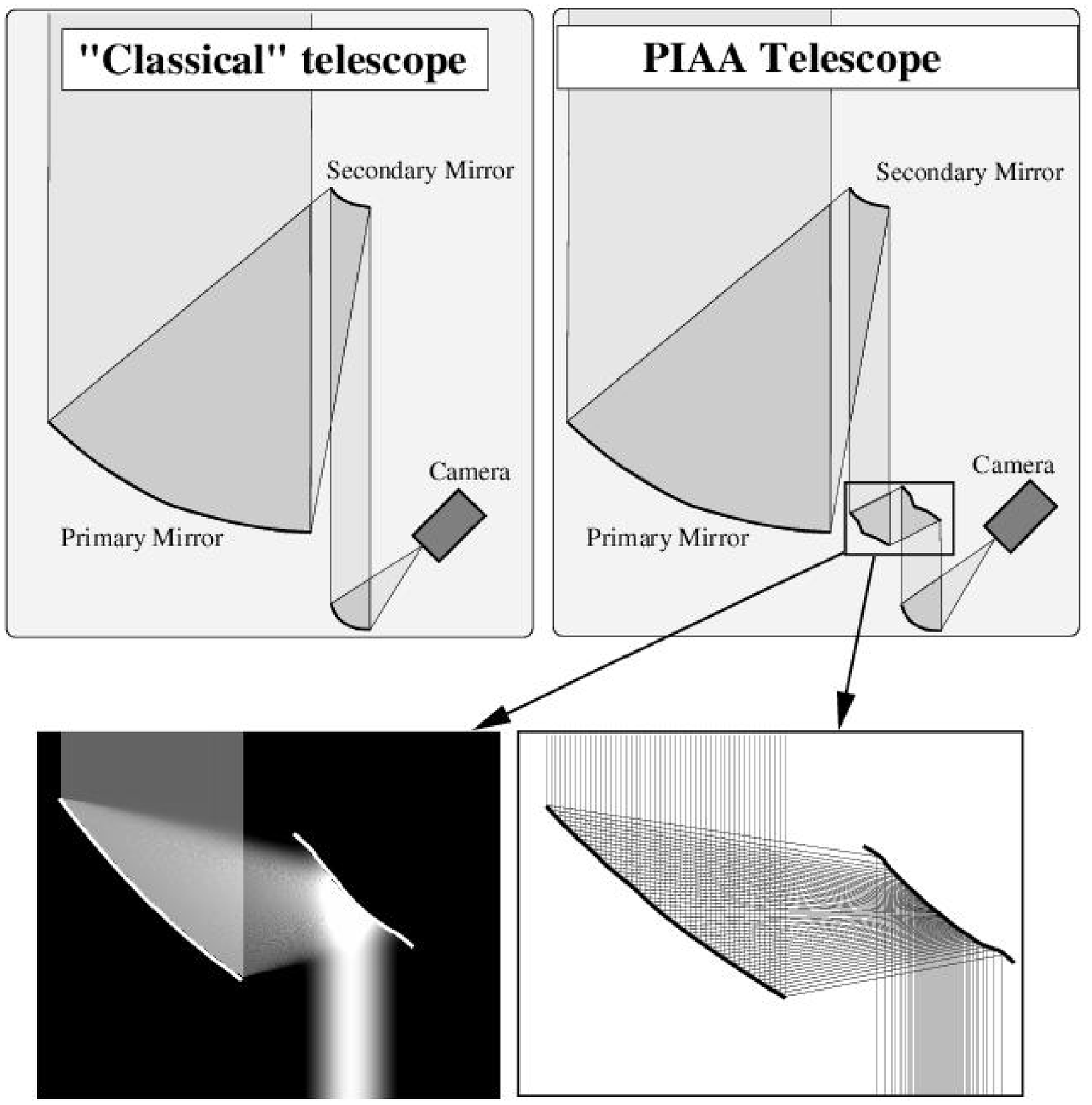}
\caption{\label{fig:piaa_principle} Schematic representation of a PIAA telescope.}
\end{figure}

As shown in figure \ref{fig:piaa_principle}, PIAA \cite{guyo03} achieves an apodization of the telescope pupil by geometric redistribution of the light. Since no light is lost, the sensitivity and angular resolution of the original telescope are preserved, and high PSF contrasts ($10^{10}$) can be achieved within $1.5 \lambda/d$ of the optical axis.


In this paper, we propose a coronagraph design based on the PIAA concept for direct imaging of extrasolar terrestrial planets. The Phase-Induced Amplitude Apodization Coronagraph (PIAAC) principle is presented in \S\ref{sec:piaac}. In \S\ref{sec:propapo}, the choice of the optimal apodization function is discussed. The field of view of the PIAAC is estimated with a raytracing model in \S\ref{sec:fov}. The performance of a PIAAC system for extrasolar terrestrial planet imaging is discussed in \S\ref{sec:sysperf}.

\section{The Phase-Induced Amplitude Apodization concepts}
\label{sec:piaac}

\subsection{Optics shapes for PIAA}
\begin{figure}[htb]
\plotone{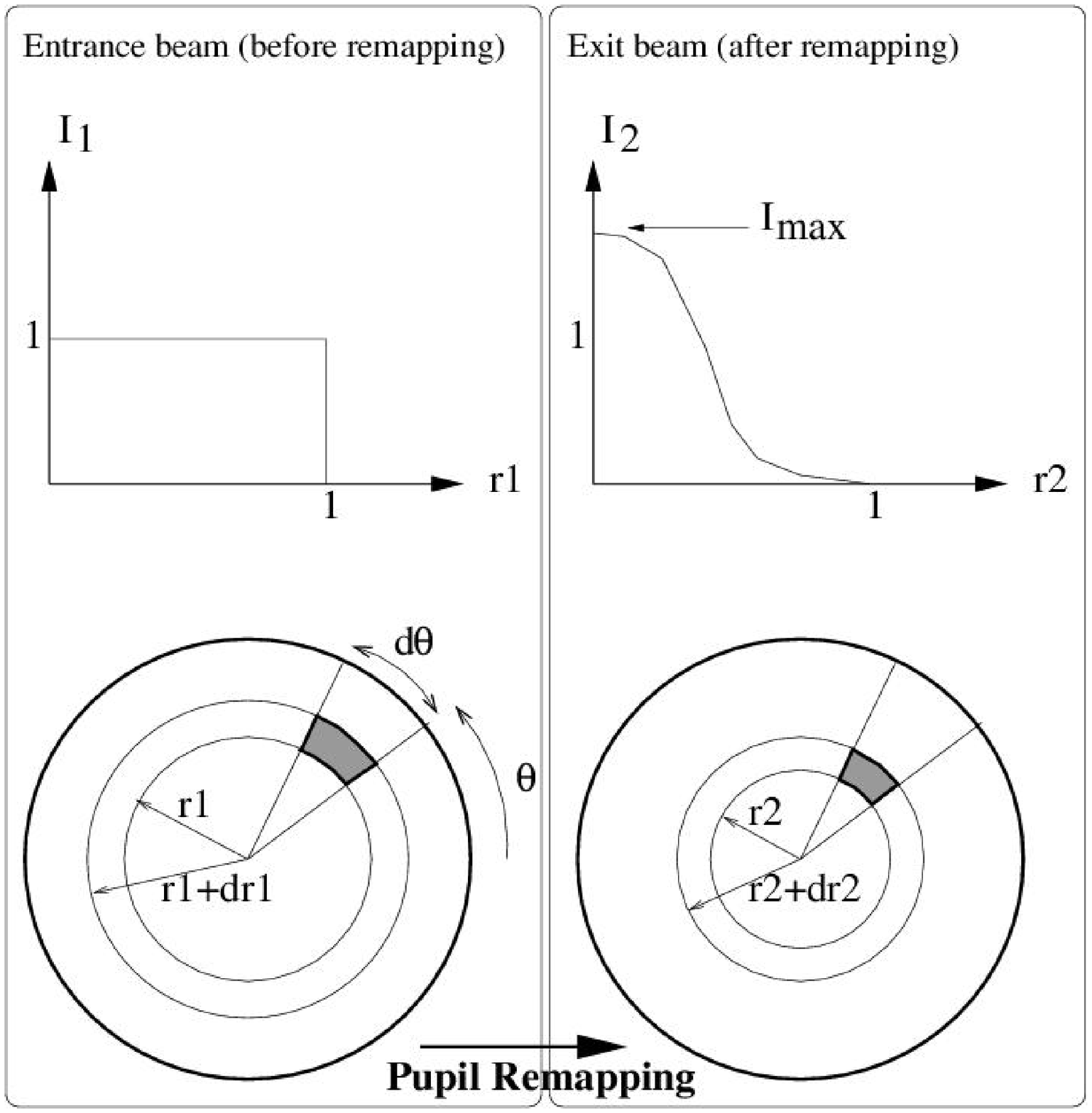}
\caption{\label{fig:remapping} Notations adopted in the paper for remapping of a pupil element in polar coordinates.}
\end{figure}

In this work, we only consider pupil remapping from a circular-symmetric intensity function to a circular-symmetric function. The surface brightness profiles of the entrance and exit beams are noted $I_1(r1)$ and $I_2(r2)$ respectively, as shown in Figure \ref{fig:remapping}. The remapping function $f$ is such that, for any value $r1$, the total flux within the radius $r1$ of the entrance beam is equal to the total flux within the radius $r2=f(r1)$ of the exit beam. The remapping function thus gives a correspondence between the radii in the entrance and exit beam.

For any desired pupil apodization, the shapes (z coordinate along the optical axis of the system) $M_1$ and $M_2$ of the mirrors for an afocal on-axis optical configuration are given by a differential equation \cite{guyo03}:
\begin{equation}
\label{equ:optshapes}
\frac{d\:M_1}{d\:r1} = \frac{d\:M_2}{d\:r2} = \sqrt{1+\left(\frac{M_2-M_1}{r1-r2}\right)^2}-\frac{M_2-M_1}{r1-r2}.
\end{equation}
The first equality in this equation guaranties that no phase aberration is introduced by the system for an on-axis point source, and the right term insures that the beam is properly apodized. This equation can also be written independently for each optical element \cite{trau03}. The optics shapes for off-axis afocal remapping with a non-circular symmetric beam can be obtained by solving a similar differential equation in 2 dimensions:
\begin{equation}
\label{equ:optshapes1x}
\frac{d\:M_1}{d\:x1} = \frac{d\:M_2}{d\:x2} = \sqrt{1+\left(\frac{M_2-M_1}{x1-x2}\right)^2}-\frac{M_2-M_1}{x1-x2}
\end{equation}
\begin{equation}
\label{equ:optshapes1y}
\frac{d\:M_1}{d\:y1} = \frac{d\:M_2}{d\:y2} = \sqrt{1+\left(\frac{M_2-M_1}{y1-y2}\right)^2}-\frac{M_2-M_1}{y1-y2}
\end{equation}
where $x1,y1$ and $x2,y2$ are the coordinates of a light ray in the entrance and exit collimated beams.


The optics shapes obtained are aspheric, and can be difficult to polish to a high level of accuracy. The outer edge of the first mirror (M1) is especially challenging to polish, because it has a small (and rapidly changing) radius of curvature: the light of the incoming pupil which is in the last percent or so of the radius needs to be spread over a large area in the exit pupil. 

Several solutions exist to relax the requirements on the accuracy of the optics surfaces, and can be combined and can be employed independently or in combination:
\begin{itemize} 
\item{{\bf Using one or several deformable mirrors (DM).} Errors on the surface of the PIAA optics produce both phase and amplitude errors in the exit beam. Using a single DM to restore the correct phase relaxes the surface accuracy requirement of the PIAA optics by a factor 100 approximately. With more than one DM, amplitude errors can also be corrected and the system can tolerate even larger surface errors.}
\item{{\bf Using more than 2 mirrors to remap the pupil.} Since the most challenging feature of the optics is a sharp bend at the outer edge of the first mirror, it is possible to design a system which performs the pupil remapping in 2 steps (3 mirrors) or more. In each step, the apodization to be performed is milder, and so is the bend at the outer edge of the mirrors.}
\item{{\bf Combining PIAA and classical apodization.} Another solution to make the ``bend'' milder is to achieve a mild apodization with PIAA, and then to use a classical apodization mask to further reduce the intensity of the edges of the apodized pupil. Since this mask would only affect the faint outer edges of the remapped pupil, it would offer the benefits of apodization without significant accompanying throughput and resolution losses.}
\end{itemize} 
Since all 3 options have a minimal negative impact on the performance of the PIAA system, the results presented in this paper also apply to PIAA systems using one or several of the above options.

\subsection{The Phase-Induced Amplitude Apodization Imager (PIAAI)}
\label{sec:piaai}
\begin{figure*}[htb]
\includegraphics[scale=0.85]{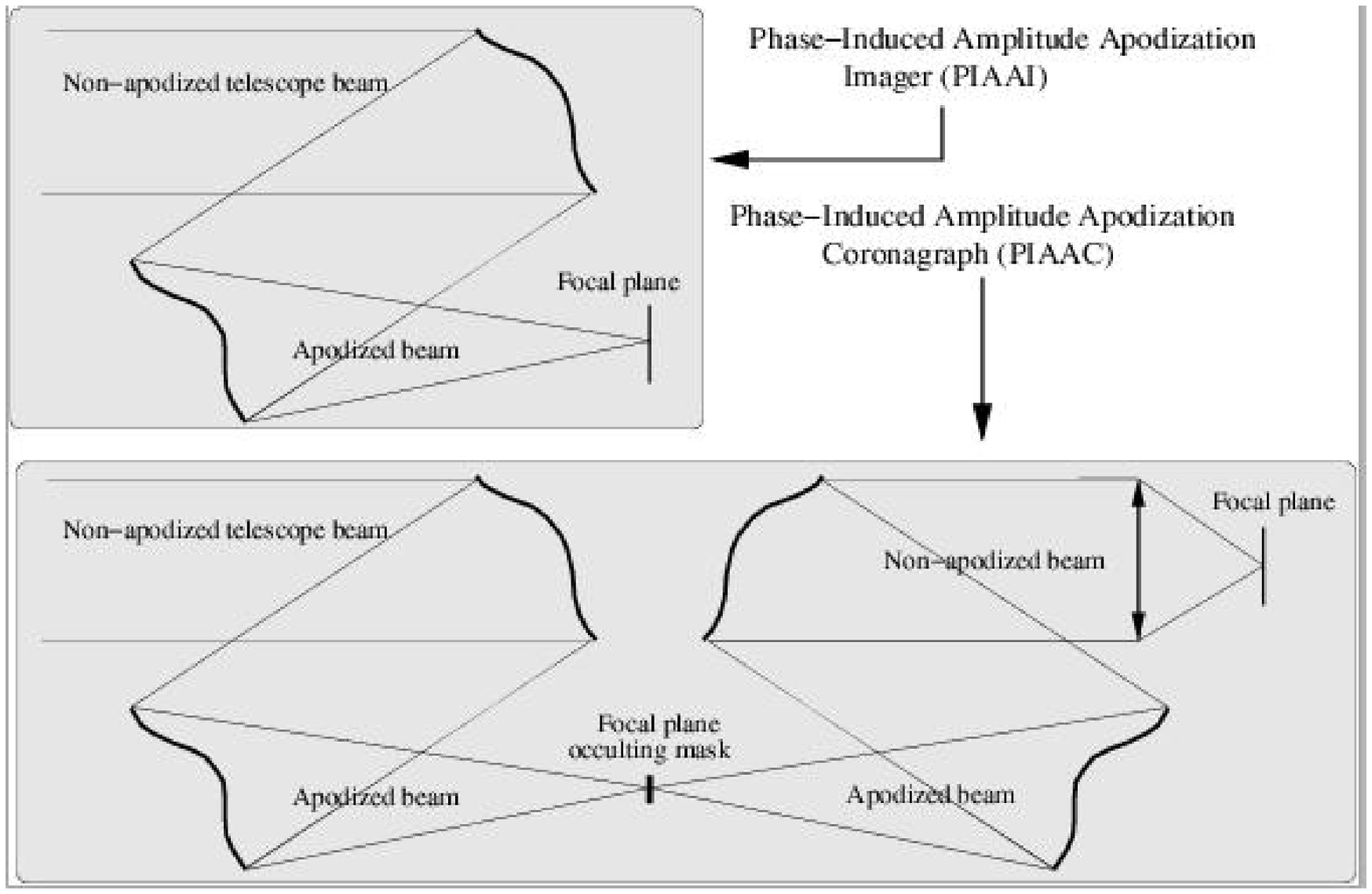}
\caption{\label{fig:PRI_PRC} Schematic representation of the Phase-Induced Amplitude Apodization Imager (PIAAI) and Phase-Induced Amplitude Apodization Coronagraph (PIAAC).}
\end{figure*}
The simplest implementation of pupil remapping on a telescope is the Phase-Induced Amplitude Apodization Imager (PIAAI), shown in fig \ref{fig:PRI_PRC} (top): the pupil is apodized by 2 mirrors, and a focal plane image is then acquired. The PIAAI is very efficient for exoplanet imaging \cite{guyo03} and offers a combination of advantages unique among high dynamical range imaging systems:
\begin{itemize}
\item{{\bf High contrast PSF}. The apodization profile can be chosen to yield PSF contrasts better than $10^{10}$.}
\item{{\bf 100 \% throughput}. Apodization does not remove light from the system: the flux sensitivity and full angular resolution of the telescope are preserved.}
\item{{\bf Small inner working angle}. A $10^{10}$ PSF contrast can be achieved at less than $2 \lambda/d$ from the optical axis.}
\item{{\bf Achromaticity}. Since the PIAAI uses mirrors and geometrical optics to perform the apodization, it is achromatic.}
\item{{\bf Low sensitivity to pointing errors}. The PIAAI behaves like a classical imaging telescope: pointing errors, or angular size of the central source, will slightly broaden the PSF core without increasing the light level in the PSF wings.}
\end{itemize}

\begin{figure}[htb]
\plotone{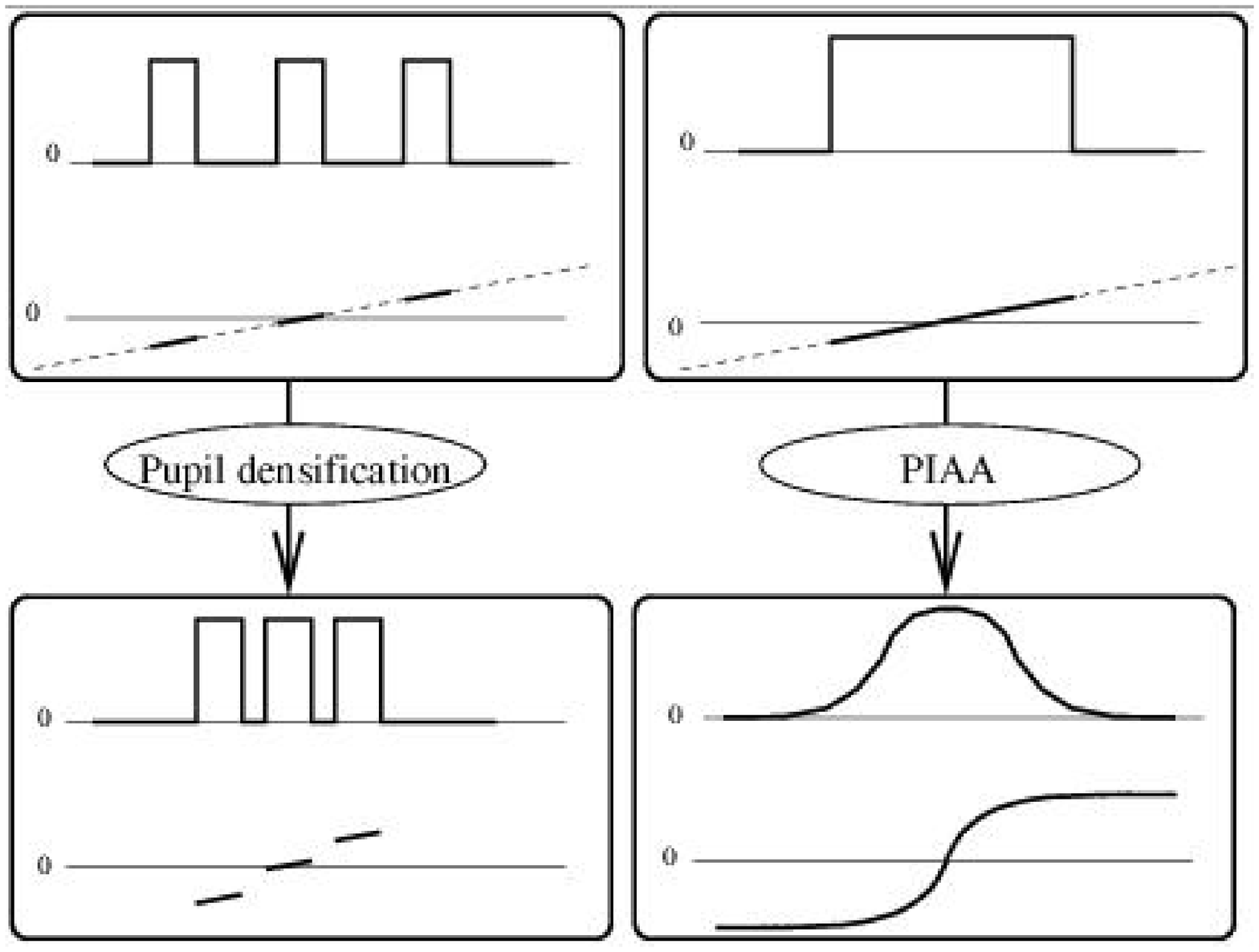}
\caption{\label{fig:pupdensphase} Effect of pupil remapping on the intensity (top graph in each square) and phase (bottom graph in each square) for an off-axis source. For both interferometers (left) and single pupil telescopes (right), the wavefront after remapping is no longer a flat tilted plane.}
\end{figure}

However, geometrical redistribution of the light in the pupil plane comes at the cost of a reduced isoplanatic field of view. As shown in Figure \ref{fig:pupdensphase}, it is impossible for a pupil remapping system to produce a pure tilted wavefront for off-axis sources. This law applies to both interferometers \cite{labe96} and monolithic pupils \cite{guyo03,soum03,trau03}, and prevents the PIAAI from directly forming good quality wide field images.

Since the apodized pupil is created by ``concentrating'' light in the central part of the pupil and ``diluting'' light in the outer parts of the pupil, the phase slope is amplified in the central part while it is decreased toward the edges. The image of an off-axis source is therefore distorted: the central part of the pupil (where the phase slope is the strongest) creates a diffraction peak far from the optical axis, while the fainter outer parts of the pupil (low phase slope) create a PSF ``tail'' pointed towards the optical axis (this correspondence between phase slope in the remapped pupil and structures in the off-axis PSF is illustrated in Figure \ref{fig:offaxis}, and will be detailed in \S\ref{ssec:optprof}). 

At small angular distances, the PSF aberrations are moderate, and do not have a strong impact on the detection sensitivity for extrasolar terrestrial planet imaging at less than about $5 \lambda/d$. However, at larger separations, the light of the planet is spread over a wide area, and the amount of background light (from the central star, the zodiacal light, the exozodiacal light and possibly from images of other sources in the field) mixed with the signal reduces sensitivity. In the next section, we demonstrate that this limitation can be overcome at a minimal cost.

\subsection{The Phase-Induced Amplitude Apodization Coronagraph (PIAAC)}
Since the PIAAI concentrates the central star light in a single diffraction peak, with very little energy outside this peak, an occulting mask may be placed in the PIAAI's focal plane to efficiently remove star light. The light of an off-axis source is not blocked by the mask, and a coronagraphic effect is therefore produced. More importantly, the beam may be de-apodized after this occulting mask to remove the off-axis wavefront distortion introduced by the pupil remapping. This design, noted PIAAC (Phase-Induced Amplitude Apodization Coronagraph) is shown in Figure \ref{fig:PRI_PRC} (bottom). The PIAA technique is then only used to create an intermediate focal plane in which the light of the central star can be efficiently masked. If the occulting mask is removed, or if the source is sufficiently off-axis, the effects of the 2 sets of optics cancel one another, and an Airy pattern is obtained in the focal plane. The same solution had been previously proposed \cite{guyo02} to restore a wide field of view in a coronagraphic interferometer.

The PIAAC offers several improvements over the PIAAI :
\begin{itemize}
\item{{\bf Wider field of view.} The off-axis PSF aberrations of the PIAAI limit the usable field of view. This effect is mitigated by the corrective optics included in the PIAA, allowing a FOV of about 100 $\lambda/d$ in radius.}
\item{{\bf Removal of the central star's flux.} The PIAAI concentrates the stellar flux in one single diffraction peak, but does not remove it from the image. Since the PIAAC efficiently blocks most of the stellar flux, a low dynamic range focal plane detector array can be used at the reimaged focus.}
\item{{\bf Smaller inner working distance.} The PIAAC is very efficient at removing the light of the central source and transmitting light of a companion, even if the 2 PSFs partially overlap in the first focal plane. In the PIAAI, the separation of 2 overlapping PSFs is less efficient because it relies on the focal plane detector, which has to overcome high dynamic range and limited pixel scale.}
\item{{\bf Sharper PSF.} The off-axis PSF of the PIAAC is as small as the telescope's diffraction allows ($\approx \lambda/d$), even at small distance from the optical axis. A minimal amount of background signal (thermal background, zodi and exo-zodi light) is therefore mixed with the companion's image.}
\end{itemize}

Since the additional optics required to build the PIAAC are after the occulting mask (where most of the central star light has been removed), they do not need to be of high optical quality and their alignment is not as critical as the coronagraphic optics. The advantages offered by the PIAAC over the PIAAI therefore come at a small additional cost.

\section{Apodization profile and inner working angle}
\label{sec:propapo}
\subsection{Definition of the Inner working angle (IWA)}
In a classical apodized pupil imager, the apodization function is chosen to (1) minimize the inner working angle (smallest angular separation at which a companion, of a given contrast with the central source, can be ``easily'' detected), (2) maximize throughput, (3) maximize the outer working angle and (4) maximize the fraction of the field of view suitable for high contrast detection. The PIAAC enjoys 100\% throughput, large outer working angle (the apodization is continuous), and full usable field of view. The choice of the apodization function is therefore only driven by the inner working angle, which shall be used as a metric of the PIAAC apodization function performance.

The inner working angle is a function of the contrast level to be achieved, and can be noted $IWA_C$, where $C$ is the 10-base logarithm of the contrast. For direct terrestrial exoplanet imaging, a suitable PSF contrast level is $10^{10}$, and we shall therefore use $IWA_{10}$ as a metric for how close to the optical axis detection of terrestrial exoplanet is possible.

 In an imaging system with a translation-invariant PSF, measuring the $IWA_{10}$ is straightforward: it is the angle from the optical axis at and beyond which the PSF profile is below $10^{-10}$ of its peak surface brightness.

This definition is however not suitable for the PIAAC (and other coronagraphs), since, at the contrast level considered, no light of the central star is left in the focal plane. In a classical coronagraph, the throughput is a function of the angular separation with the optical axis, and the $IWA$ shall therefore be defined by imposing a minimum required coronagraphic throughput $T$. For easy comparison between coronagraphs and apodized pupil imagers, this threshold value should be chosen to be approximately equal to the throughput of a classical apodizer designed to reach similar contrast level at small $IWA$ (3 to 4 $\lambda/d$). In this work, we adopt $T=0.1$ as the threshold transmission to define the IWA.

\subsection{Focal plane scale}
By analogy with pupil densification of interferometric arrays \cite{labe96}, there are two equivalent representations of pupil apodization with a PIAA system:
\begin{itemize}
\item{(1) The apodized pupil is obtained by concentrating flux in the inner part of the pupil, but the pupil diameter is conserved. At the center of the pupil, the phase slope of an off-axis source is amplified by the remapping.}
\item{(2) The pupil surface brightness is conserved at the pupil's center, but the light is spread over a wider area in the apodized pupil: the apodized pupil is larger than the entrance pupil. At the center of the pupil, the phase slope of an off-axis source is identical in the entrance and exit (apodized) pupil of the PIAA system.}
\end{itemize}
With representation (2), the focal scale, if measured at the center of the pupil (``principal ray''), is preserved by the pupil remapping, which is convenient for defining a focal scale in the first focal plane of the PIAAC. This definition is however not exact since the phase slope of an off-axis source is not constant in the apodized pupil. We have chosen to adopt representation (1) in this work to simplify notations, which gives us 2 options for defining the focal scale :
\begin{itemize}
\item{{\bf Ignoring the phase slope amplification effect.} The focal scale is measured as if the apodized beam were obtained by classical apodization and would preserve the phase slope of off-axis sources. In a system of focal length $F$, $F (\lambda/d)_0$ is the physical distance in the focal plane that would separate the images of two sources separated by $\lambda/d$ in the sky if the apodized beam were obtained by classical apodization. In this paper, we omit $F$ and use $(\lambda/d)_0$ as a unit of distance in the focal plane. This unit does not correspond to an angular distance on the sky (because PIAA does not preserve the phase slope of off-axis sources), but can be easily obtained mathematically through Fourier transform of a remapped pupil.}
\item{{\bf Adopting a physically meaningful focal scale from the principal ray of the system.} The focal scale can be measured in the focal plane from the displacement of the principal ray (the light ray at the center of the pupil) of the system for a source in the sky at a known angular distance from the optical axis. We denote $F (\lambda/d)_s$ (in the rest of the paper, we omit $F$) this displacement in the focal plane for a source which is moved by $\lambda/d$ in the sky. Considering only the principal ray does not however take into account the change of PSF shape with separation.}
\end{itemize} 

With conventional imaging, the image scale is unambiguously determined by the mapping between imaged points. In PIAA the ambiguity of image scale can be associated with the distorted PSF (which in PIAAI is a feature, not an aberration). Lacking a well-defined image center for the PSF, the scale is not simply determined. The scale $(\lambda/d)_s$ corresponds to what appears as the offset ``core'' of the PSF, corresponding approximately to the intensity peak, not a weighted mean.

As will be demonstrated in the next section, with the notations adopted in this paper (see Figure \ref{fig:remapping}), for a ``smooth'' apodization profile,
\begin{equation}
(\lambda/d)_s = \sqrt{I_{max}} (\lambda/d)_0.
\end{equation}
The ratio between the 2 scales corresponds to the ``Angular Magnification'' factor defined  by Traub \& Vanderbei (2003), and can be a source of confusion when attempting to measure the IWA in a PIAAI system. We therefore use both $(\lambda/d)_0$ and $(\lambda/d)_s$ in this paper to measure size of features in the intermediate focal plane of the PIAAC. $(\lambda/d)_s$ is preferentially used to measure the occulting mask size since it is closely related to the IWA of the system.

Fortunately, in the PIAAC, the output pupil (after the second pupil remapping) of the system is similar to the pupil of a classical telescope: the phase slope of an off-axis source is constant across the pupil and the PSF is invariant by translation. There is no focal scale problem in the final focal plane of the PIAAC, and the classical $\lambda/d$ scale is then used.

\subsection{Optimizing the apodization profile}
\label{ssec:optprof}
In the PIAAC, the focal plane mask blocks the central diffraction peak, and the transmission of the coronagraph for an off-axis source is approximately the fraction of the off-axis PSF flux which ``misses'' this mask. 

While a rigorous diffraction simulation of the coronagraph is needed to compute the throughput for off-axis sources, we use in this section a diffraction/raytracing approximation to show that the optimization problem is quite different than for the choice of an apodization function in classical apodization. The notations adopted in the pupil plane are shown in Figure \ref{fig:remapping}, and we denote by $f$, such that $r2=f(r1)$, the remapping function, and $f'(r1) = df(r1)/dr1$ its derivative.

\begin{figure}[htb]
\includegraphics[scale=0.75]{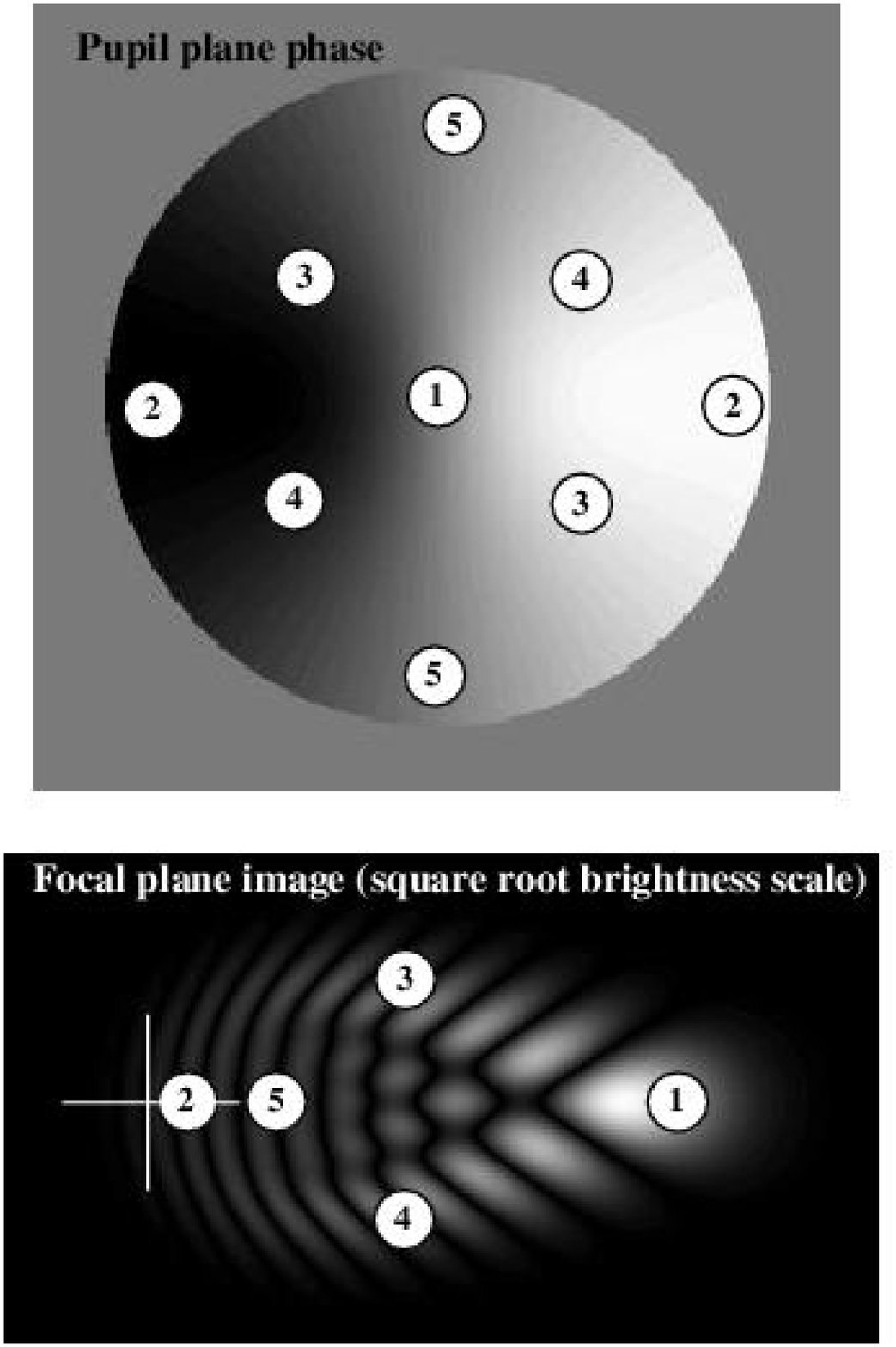}
\caption{\label{fig:offaxis} Pupil phase map (top) and PSF (bottom) for an off-axis source at 10 $\lambda$/d from the optical axis. The position of the optical axis in the focal plane is shown as a white cross in the PSF image. Numbers from 1 to 5 are used to show the correspondence between regions of the pupil and features in the off-axis PSF.}
\end{figure}

The size of the central diffraction peak in the image of an on-axis source in the first focal plane is the basis for defining the size of the occulting mask. We denote by $\alpha_0$ the radius of this mask (in units of $(\lambda/d)_0$), which is equivalent to the classical definition of the inner working distance for a classically apodized pupil imaging telescope. The phase in the entrance pupil of the PIAAC is a flat tilted wavefront for an off-axis source. The slope of this wavefront is noted $Sl1$ (normalized such that it is equal to 1 for an angular distance of $\lambda/d$ on the sky), and is aligned with the $x$ axis ($\theta = 0$ in polar coordinates).
As can be seen in Figure \ref{fig:remapping}, the remapping amplifies the radial slope (equal to $Sl1 \cos(\theta)$) by $1/f'(r1)$ and the angular slope (equal to $Sl1 \sin(\theta)$) by $r1/f(r1)$. Therefore, after remapping, for the pupil element shown in Figure \ref{fig:remapping}, the slopes along the $x$ and $y$ axis are:
\begin{equation}
\label{equ:sl1x}
Sl2_x(r1,\theta) = Sl1 \left( \frac{\cos^2(\theta)}{f'(r1)} + \sin^2(\theta) \frac{r1}{f(r1)}\right)
\end{equation}
\begin{equation}
\label{equ:sl1y}
Sl2_y(r1,\theta) = Sl1 \cos(\theta) \sin(\theta) \left( \frac{1}{f'(r1)} - \frac{r1}{f(r1)} \right).
\end{equation}
According to equ. \ref{equ:sl1y}, the remapping introduces a wavefront slope along the axis perpendicular to the direction defined by the position of the off-axis source. This effect can be seen in the PIAAI images as a broadening of the off-axis PSF tails along the radial direction (Figure \ref{fig:offaxis}). 

\begin{figure*}[htb]
\includegraphics[scale=0.85]{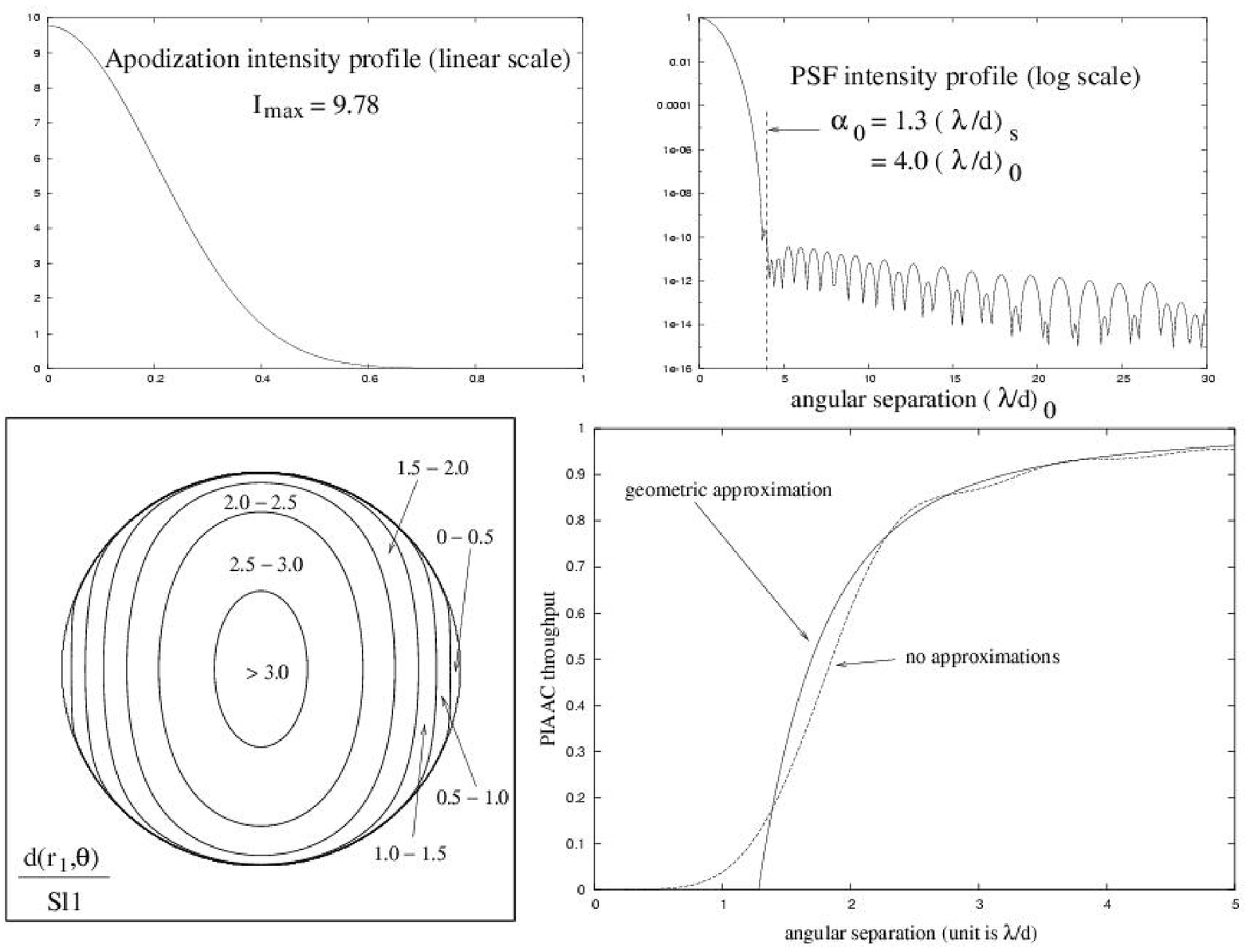}
\caption{\label{fig:opt} Apodization profile (top left) obtained with the optimization scheme described in the text. The corresponding PSF intensity profile (top right) has an intrinsic inner working angle of 4 $(\lambda/d)_0$ (1.3 $(\lambda/d)_s$ on the sky). For this profile, the factor by which the pupil phase slope is amplified is shown (lower left) on the entrance pupil of the system. The throughput computed from this map (lower right, ``geometric approximation''  curve) agrees well with the true throughput of the PIAA coronagraph.}
\end{figure*}

For each pupil element, the corresponding light ray will intersect the focal plane at a distance $d(r1,\theta)$ from the center of the mask:
\begin{equation}
d(r1,\theta) = \sqrt{\left(Sl2_x(r1,\theta)\right)^2 + \left(Sl2_y(r1,\theta)\right)^2}
\end{equation}
\begin{equation}
\label{equ:dr0theta}
d(r1,\theta) = Sl1 \sqrt{\left(\frac{cos(\theta)}{f'(r1)}\right)^2 + \left(\frac{sin(\theta) r1}{f(r1)}\right)^2}
\end{equation}
Since $Sl1$ is normalized to the slope of a source at a separation equal to $\lambda/d$, $d(r1,\theta)$ is expressed in units of $(\lambda/d)_0$. $d(r1,\theta)/Sl1$ represents the factor by which the phase slope in the entrance pupil has been multiplied by the pupil remapping. As shown in Figure \ref{fig:opt} (bottom left), it is maximum at the center of the pupil (where the light needs to be concentrated) and minimum at the edges of the pupil (where light needs to be diluted to produce the required apodization profile).

The occulting focal plane mask acts as a ``wavefront slope high pass filter'', as it will block the light of pupil elements for which $d(r1,\theta)$ is smaller than $\alpha_0$. The transmission of the coronagraph for an off-axis source is therefore obtained by the following integration over the entrance pupil ($0<r1<1,\: 0<\theta<2\pi$) of the coronagraph:
\begin{equation}
\label{equ:Talpha}
T = \frac{1}{\pi} \int_{r1,\theta}{ H\left(\frac{d(r1,\theta)}{\alpha_0}\right) r1 \: dr1 \: d\theta}
\end{equation}
where $H(x)$ is equal to 1 if $x>1$, and equal to 0 is $x<1$.

If the pupil remapping preserves the original pupil ($f(r1)=r1$, $f'(r1)=1$, $Sl2_x=Sl1$, $Sl2_y=0$) $d(r1,\theta) = Sl1$ and $T$ is therefore equal to 1 if $Sl1>\alpha_0$ and to 0 if $Sl1<\alpha_0$: the inner working distance is then equal to $\alpha_0$, just as for a classical apodizer.

The example of $d(r1,\theta)/Sl1$ given in Figure \ref{fig:opt} (bottom left) shows that, as a point source is moved further away from the optical axis, the central zone of the entrance pupil (which also corresponds the the central zone of the exit pupil) is first transmitted by the coronagraph. At large separations, the full pupil is transmitted, and the PIAAC throughput is close to $100\%$, as shown in Figure \ref{fig:opt} (bottom right). There is a very good agreement between the true throughput of the PIAAC (obtained by taking into account diffraction) and the estimation obtained by equ. \ref{equ:Talpha}.

For an apodization profile which has a maximum surface brightness $I_{max}$ (normalized to the entrance pupil surface brightness) at its center, the PIAAC will start to transmit the light of an off-axis source when the phase slope in the central part of the apodized pupil exceeds $\alpha_0$. If the apodization profile is smooth, the following approximations are valid in the central part of the pupil:
\begin{equation}
\label{equ:fr0approx}
f(r1) \approx \frac{r1}{\sqrt{I_{max}}} 
\end{equation}

\begin{equation}
\label{equ:fpr0approx}
f'(r1) \approx \frac{1}{\sqrt{I_{max}}}.
\end{equation}
From equ. \ref{equ:fr0approx}, \ref{equ:fpr0approx} and \ref{equ:dr0theta}, we can write
\begin{equation}
\label{equ:dr0thetaapprox}
d(r1,\theta) \approx Sl1 \sqrt{I_{max}}.
\end{equation}

From equ. \ref{equ:Talpha} and \ref{equ:dr0thetaapprox}, we can estimate the approximate value of the inner working angle according to our diffraction/raytracing study:
\begin{equation}
IWA [\lambda/d] \approx \frac{\alpha_0[(\lambda/d)_0]}{\sqrt{I_{max}}} \approx \alpha_0[(\lambda/d)_s]
\end{equation}
where $IWA [\lambda/d]$ is expressed in units of $\lambda/d$, and $\alpha_0[(\lambda/d)_0]$ and $\alpha_0[(\lambda/d)_s]$ are the radius of the occulting mask expressed in units of $(\lambda/d)_0$ and $(\lambda/d)_s$ respectively. This important result shows that the inner working angle is approximately equal to the ratio of the ``classical'' definition of the inner working angle of an apodizing profile to the square root of the central brightness of this profile. This factor is the ``Effective Angular Magnification'' factor obtained by numerical simulations by Traub \& Vanderbei (2003).
In a classical apodizer, $I_{max} = 1$, and $IWA = \alpha_0[(\lambda/d)_0]$, but in the PIAAC, $I_{max} > 1$ (light is concentrated in the central part of the apodized pupil), and the inner working angle can therefore be smaller than $\alpha_0[(\lambda/d)_0]$. 

To find the optimal apodization profile for the PIAAC, the ``classical'' inner working angle $\alpha_0$ needs to be minimized and the central brightness $I_{max}$ needs to me maximized: these 2 requirements are somewhat opposite, as increasing $I_{max}$ usually also increases $\alpha_0$. In the usual ``classical'' apodization profile optimization, where $I_{max}=1$, only $\alpha_0$ is to be minimized.

\subsection{Results of the optimization}
An iterative optimization algorithm was used to find an apodization profile suitable for the PIAAC. The algorithm, which is similar to the one used in a preliminary study of the PIAAI \cite{guyo03}, computes $\alpha_0$ (using Fourier optics) and the scaling factor ($\sqrt{I_{max}}$) at each iteration. A ``smoothness'' constraint is imposed on the pupil to prevent the optimization algorithm from maximizing $\sqrt{I_{max}}$ by producing a narrow peak in the apodization profile. Such a narrow peak would increase $\sqrt{I_{max}}$ without improving the IWA of the system, because it would contain a small fraction of the total light.

The profile obtained by this optimization is shown in Figure \ref{fig:opt}. The parameters of this apodization are: $\alpha_0 = 4.0 (\lambda/d)_0$, $I_{max} = 9.73$. The profile is very close to a prolate spheroidal function, and no significant performance difference was found between the apodization profile delivered by the optimization algorithm and a prolate spheroidal function of similar width. The prolate spheroidal function seems to be the optimal solution for both classical apodization and PIAA. The main benefit of the PIAA is that it does not suffer from loss of angular resolution or throughput that the prolate spheroidal function would produce in a classical apodization. Therefore, with the PIAA, an even larger region of the extended wings of the prolate spheroidal function could be included in the apodization profile without loss of performance: the scale of the apodization function becomes irrelevant.

\section{Field of view aberrations in the PIAAC}
\label{sec:fov}
\subsection{Off-axis PSFs close to the optical axis}
\begin{figure}[htb]
\includegraphics[scale=0.6]{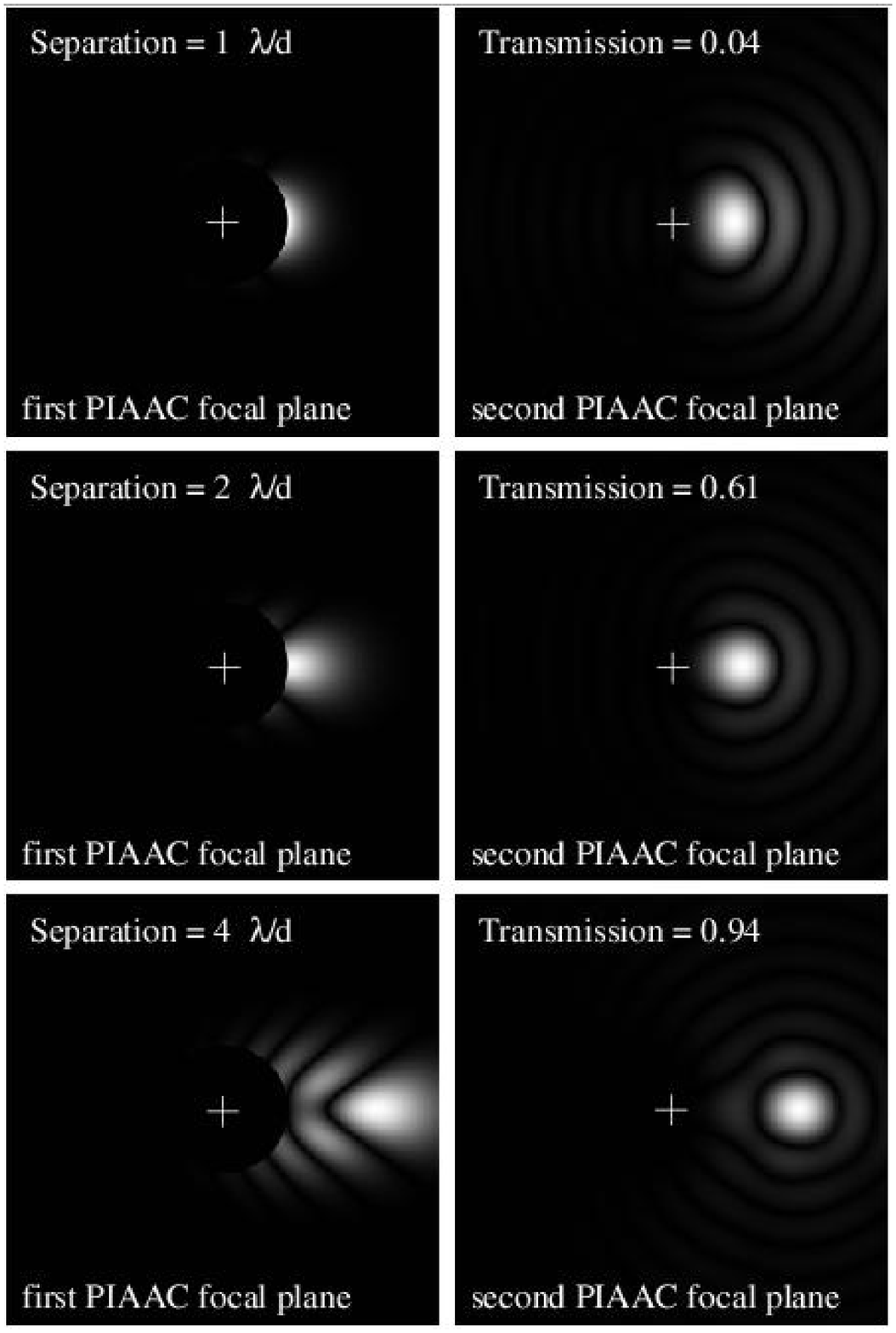}
\caption{\label{fig:piaacpsfs} Light intensity distributions in the first focal plane (immediately after the occulting mask) of the PIAAC (left) and in the second focal plane (right) for off-axis sources at 1 $\lambda/d$, 2 $\lambda/d$ and 4 $\lambda/d$ from the optical axis. The white cross indicates the position of the optical axis.}
\end{figure}
As a point source is moved away from the optical axis of the PIAAC, its image in the first focal plane will simultaneously drift outside the occulting mask and become distorted because of the field aberrations of the PIAAI (Figure \ref{fig:piaacpsfs}). In the final focal plane of the PIAAC, the image of an off-axis source far from the optical axis is the Airy pattern, and no light is blocked by the occulting mask. However, for sources at separations comparable to the IWA, the PSF shape as well as the throughput are affected by the occulting mask. Figure \ref{fig:piaacpsfs} (right) shows the shape of the PSF for off-axis sources. The PSF is elongated around the optical axis for small separations (less than $2 \lambda/d$), and rapidly becomes very close to the Airy pattern beyond $2 \lambda/d$. 

The effect of the occulting mask in the image is therefore very localized, and the PSF of sources outside the IWA is not broader than the diffraction limit of the telescope. The PIAAC efficiently removes the light of the central source while delivering a high-quality image of the rest of the field.
 
\subsection{Narrow field approximations}
We have assumed so far that the PIAA is a ``pure remapping'' system: the PIAA output pupil is obtained by remapping the phase and amplitude of the PIAA entrance pupil. As shown in fig. 2, the PIAA system is then entirely defined by the remapping function $f$ which gives the correspondence between points in the entrance and exit pupils.
This assumption is valid for an on-axis point source (the optics shapes have been designed for this case), but is only an approximation for off-axis sources. Phase and amplitude aberrations appear as the tilt angle of the incoming light rays becomes sufficient to move the impact point of light rays on the surface of the optics (``beam walk'' effect).

\begin{figure}[htb]
\plotone{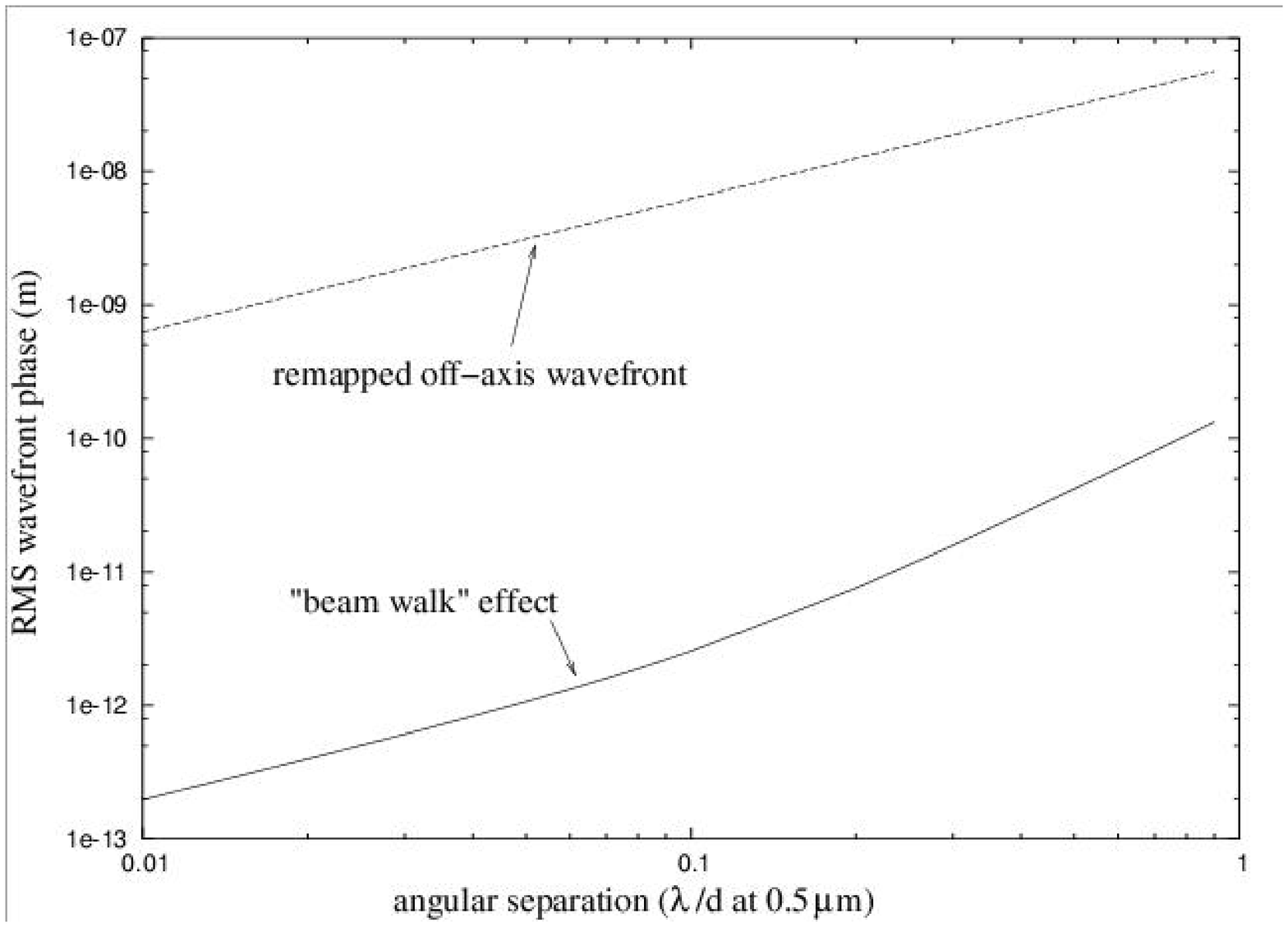}
\caption{\label{fig:beamwalk} RMS wavefront quality of the ``beam walk'' phase mode as a function of angular separation of the source to the optical axis. For comparison, the RMS wavefront quality of the remapped beam is also shown. In both cases, the wavefront was intensity-weighted prior to computation of the RMS wavefront phase.}
\end{figure}

We have measured, through numerical simulation (3D raytracing on the surface of the optics), the deviation from ``perfect'' pupil remapping in a real system. The main effect produced by beam walk is to superimpose a low-order phase mode on the wavefront exiting the PIAA system. As shown in fig. \ref{fig:beamwalk}, the RMS amplitude of this mode increases with the source's angular separation from the optical axis. For separations below $0.1 \lambda/d$, the phase aberration induced by the beam walk effect is about 3000 times smaller than the remapped tip-tilt phase map. This effect has therefore no significant impact on the sensitivity of the PIAAC to angular stellar size and pointing errors.

\subsection{Wide field of view with the PIAAC}

\begin{figure}[htb]
\plotone{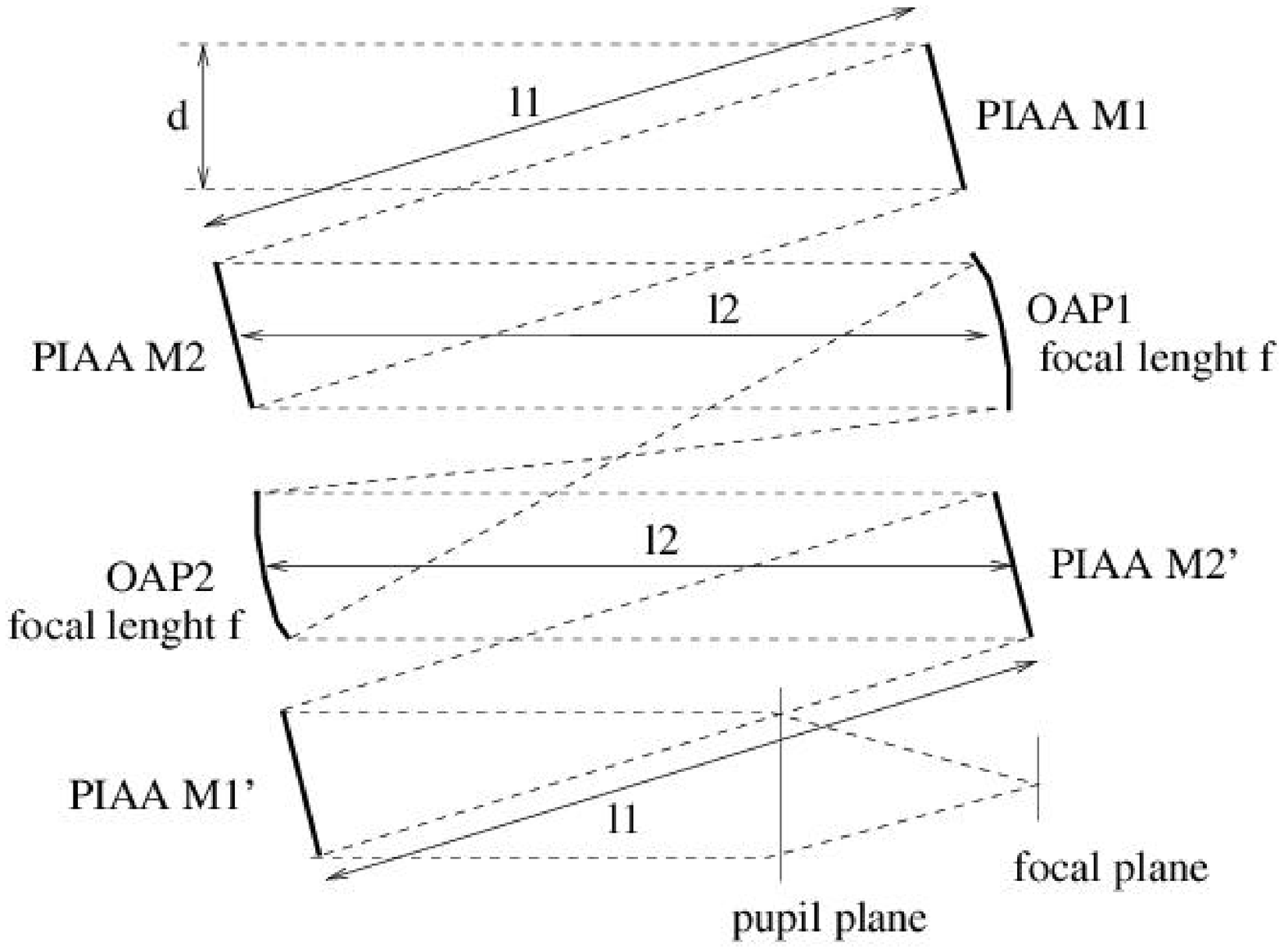}
\caption{\label{fig:raytracetest} Geometry of the PIAAC system used for the raytracing simulation. The beam diameter, $d$ is 75mm, the distance $l1$ between the PIAA mirrors is $15 \times d = 1125$mm, the focal $f$ of the OAPs is $5.33 \times d = 400$mm and the distance $l2$ between the OAPs and the PIAA mirrors was set to $2.67 \times d = 200$mm (see fig. \ref{fig:piaacaberr020}) and $9.6 \times d = 720$mm (see fig. \ref{fig:piaacaberr072}).}
\end{figure}

The PIAAC uses a second set of aspheric optics to cancel the off-axis aberrations introduced by the pupil remapping. The beam walk effect described above also limits the field of view within which the aberration induced by the first set of optics can be canceled with a second set of optics. We have used a 3D raytracing simulation to simulate this effect. Figure \ref{fig:raytracetest} shows the geometry of the PIAAC system adopted for this test. 

\begin{figure}[htb]
\plotone{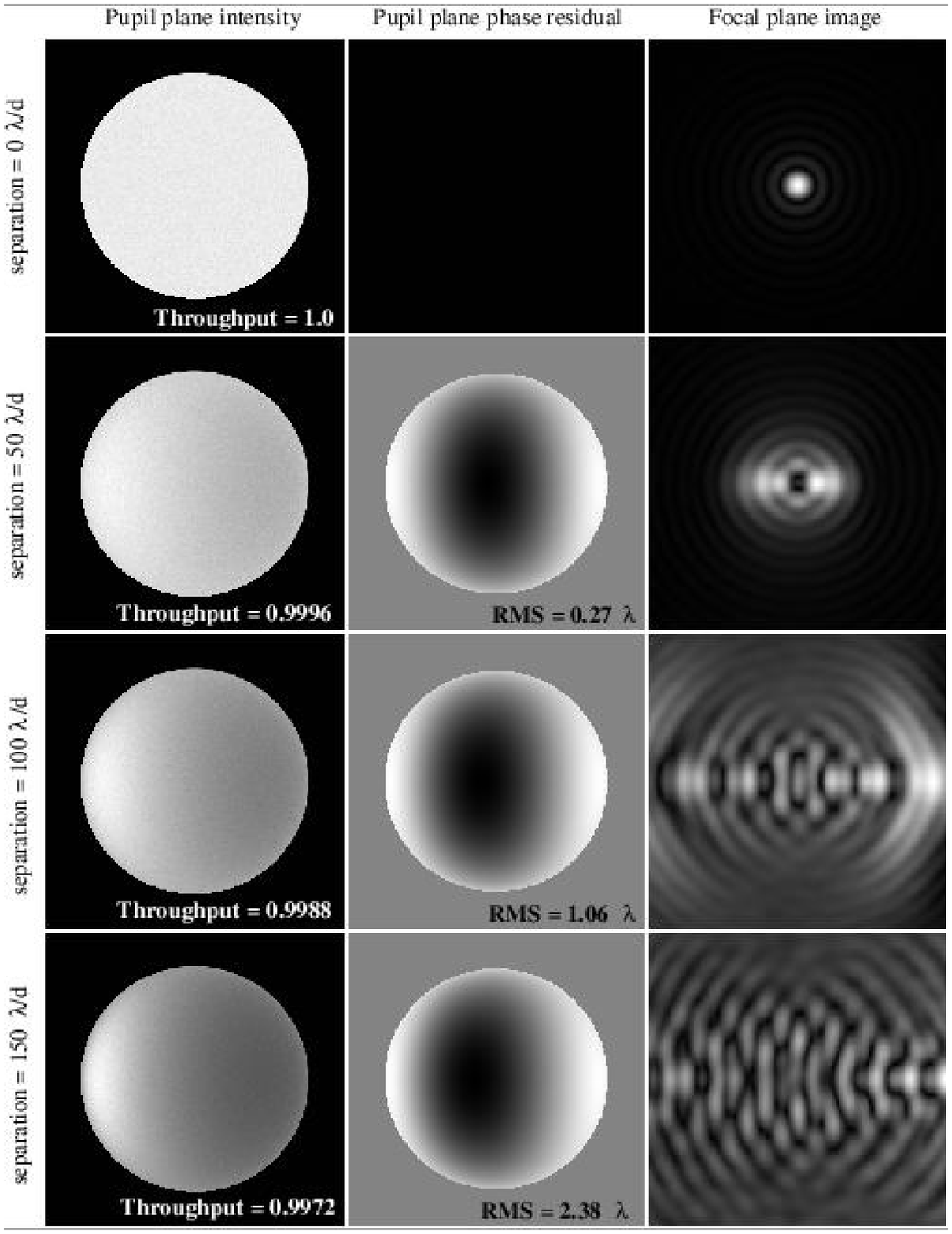}
\caption{\label{fig:piaacaberr020} Pupil plane intensity (left column), phase (center column) and corresponding PSF (right column) for off-axis sources with the PIAAC. The tip-tilt mode has been removed from the phase map to show the off-axis aberrations introduced by the PIAAC. The wavelength is $0.5 \mu$m in this monochromatic simulation.}
\end{figure}

We first tested this geometry with a distance $l2$ between the OAPs and the PIAA mirrors set to $2.67 \times d = 200$mm (see fig.\ref{fig:piaacaberr020}). The results show that the field of view under this configuration is less than $50 \lambda/d$ in radius: the beam walk effect is rapidly broadening the PSF at and outside this radial distance.

This configuration is however not optimal, and we have then adopted the distance $l2$ between the OAPs and PIAA mirrors which maximizes the field of view: $9.6 \times d = 720$mm. The field of view is then increased to more than $100 \lambda/d$ in radius. For the second set of PIAA mirrors to cancel the first set over a large field of view, the PIAA mirrors in the first set should be conjugated to the corresponding mirrors in the second set. While this cannot be achieved with the optical configuration shown in fig. \ref{fig:raytracetest} (additional mirrors would be required to conjugate M1 to M1' and M2 to M2'), one degree of freedom (value of $l2$) is sufficient to more than double the field of view of the system. With more optical elements and more degrees of freedom, a wider field of view may be achieved.

\begin{figure}[htb]
\plotone{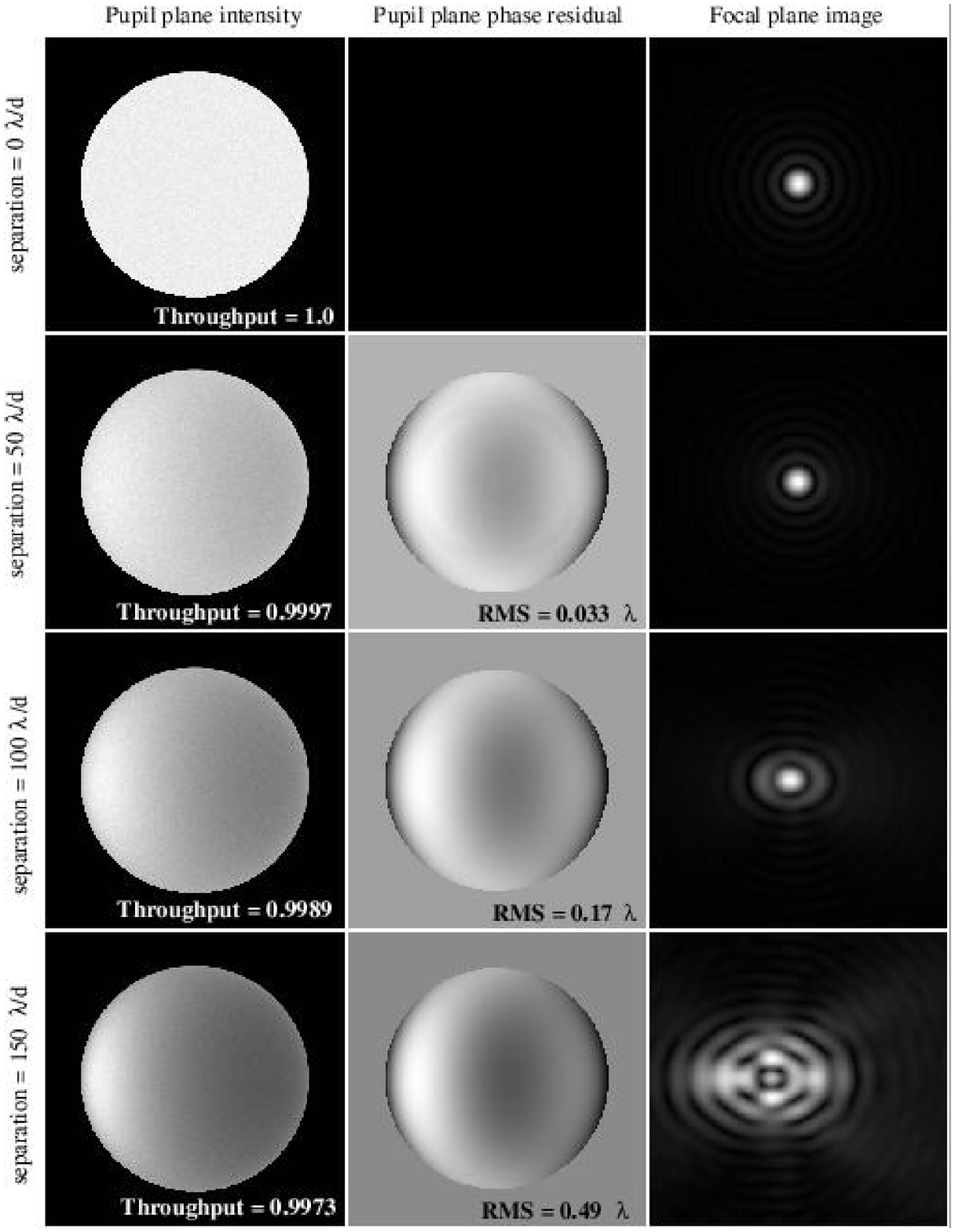}
\caption{\label{fig:piaacaberr072} Same as figure \ref{fig:piaacaberr072} with a different value for $l2$. Here, $l2$ has been chosen to maximize the image quality for off-axis sources.}
\end{figure}

\section{System performance}
\label{sec:sysperf}

\begin{figure*}[htb]
\includegraphics[scale=0.83]{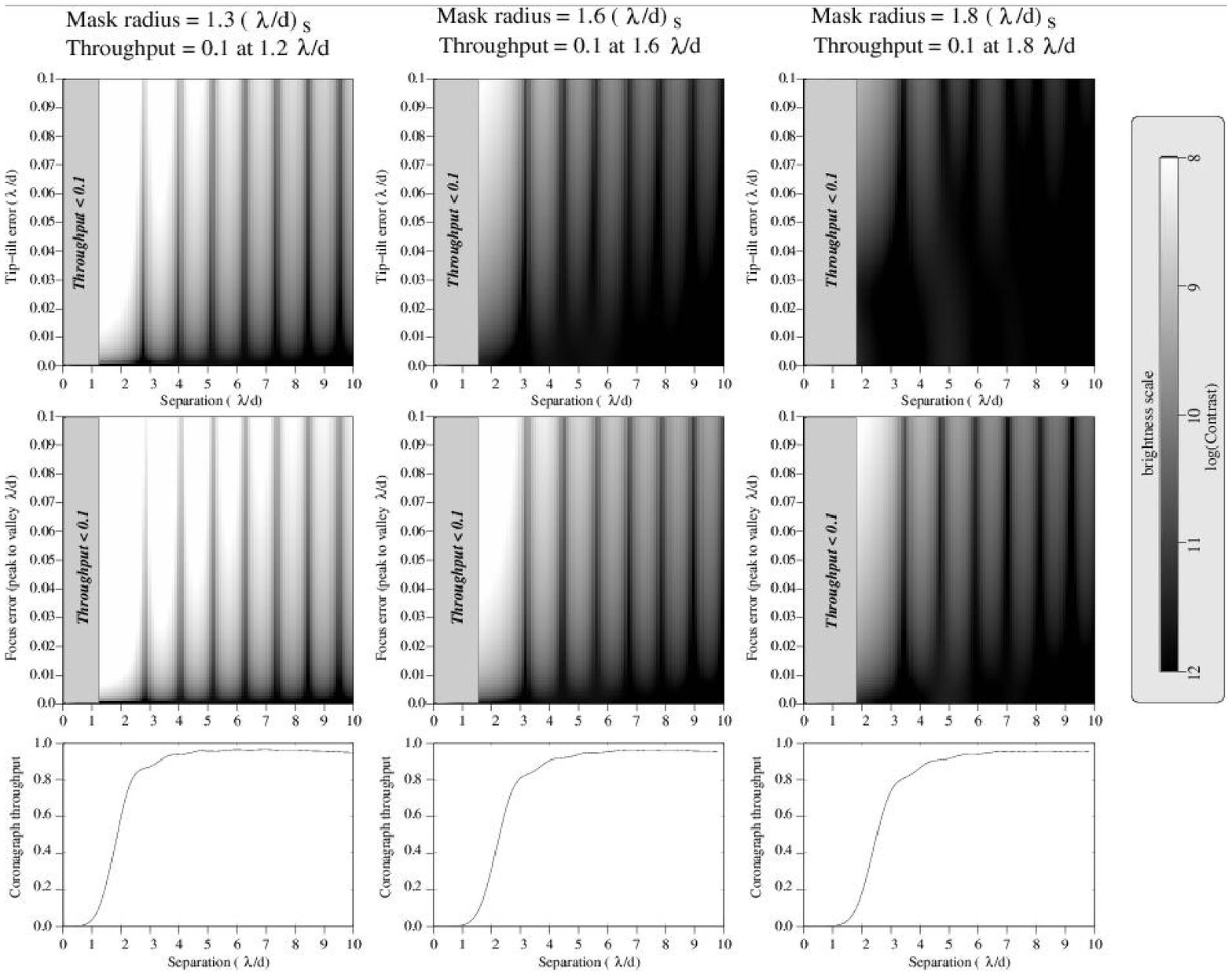}
\caption{\label{fig:errsens} Sensitivity of the PIAAC to tip-tilt and focus wavefront errors for 3 different sizes of focal plane occulting mask (mask size increases from left to right). The raw contrast of the PIAAC image is plotted for tip-tilt (top row) and focus (center row) errors as a function of companion offset (x axis) and amplitude of the wavefront error (y axis). The PIAAC throughput as a function of source offset is also shown (bottom row). The range of offset for which the PIAAC throughput is less than 10\% is shown in grey in the top and center rows.}
\end{figure*}

\subsection{Compatibility with telescope pupil shapes}
\begin{figure}[htb]
\plotone{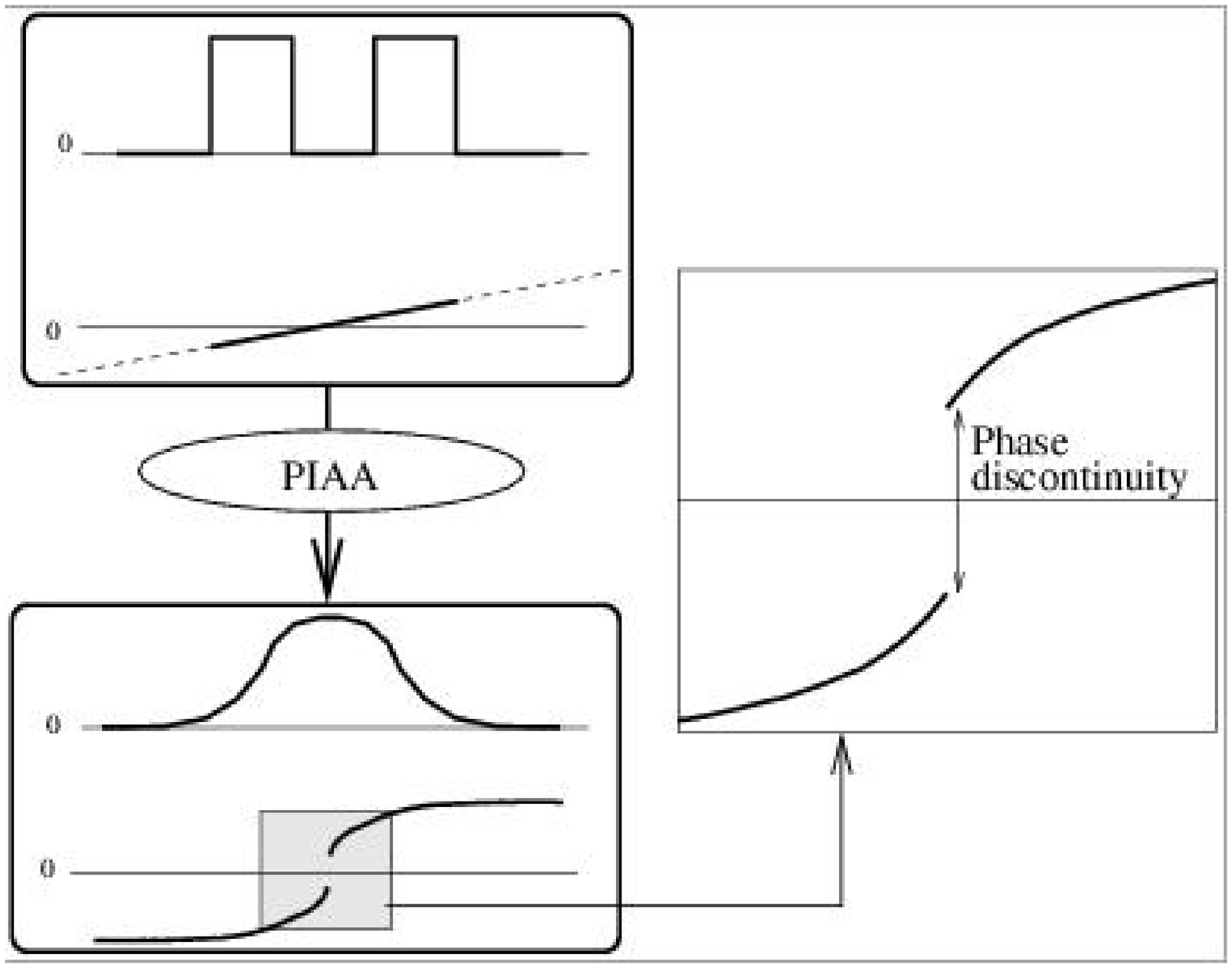}
\caption{\label{fig:centobs} Effect of a central obstruction in the remapping of an off-axis source wavefront by the PIAA. A phase discontinuity appears because of the ``filling'' of the central obstruction.}
\end{figure}
A telescope pupil with a central obstruction may be remapped into an apodized pupil without central obstruction, and a PIAAC can therefore be built for a on-axis telescope. Just as for the concept presented so far, the IWA and contrast level would be suitable for direct imaging of Earth-sized planets. However, as shown in fig. \ref{fig:centobs}, the remapping of the central obstruction produces a phase discontinuity in the wavefront of off-axis sources. At the $10^{10}$ contrast level required to detect Earth-sized planets, this effect makes the PIAAC too sensitive to pointing error and stellar diameter unless the central obstruction is made very small. The same effect imposes constraints on the thickness of the spiders vanes needed to support the secondary mirror. 

It is however possible to apodize both the outer edge and the inner edge of a pupil with a central obstruction, but the IWA for Earth-sized planet detection is then significantly larger.

Elliptical telescope pupils can easily be remapped into circular pupils (with ``cylindrical'' optics), which can themselves be remapped to be suitable for high contrast imaging. These 2 remapping steps can be combined into one, and the PIAAC concept presented in this paper can be generalized to elliptical apertures.

\subsection{Sensitivity to low order aberrations and focal plane mask size}
\label{ssec:ttsens}
Phase aberrations in the entrance beam yield a PSF which ``overfills'' the occulting mask in the first focal plane of the PIAAC. Low order aberration broadens the central diffraction peak of the PSF without significantly increasing the faint PSF wings. Increasing the size of the occulting mask therefore makes the PIAAC quite insensitive to low order aberrations, including tip-tilt and focus, as shown in fig. \ref{fig:errsens}.

\begin{figure*}[htb]
\includegraphics[scale=0.9]{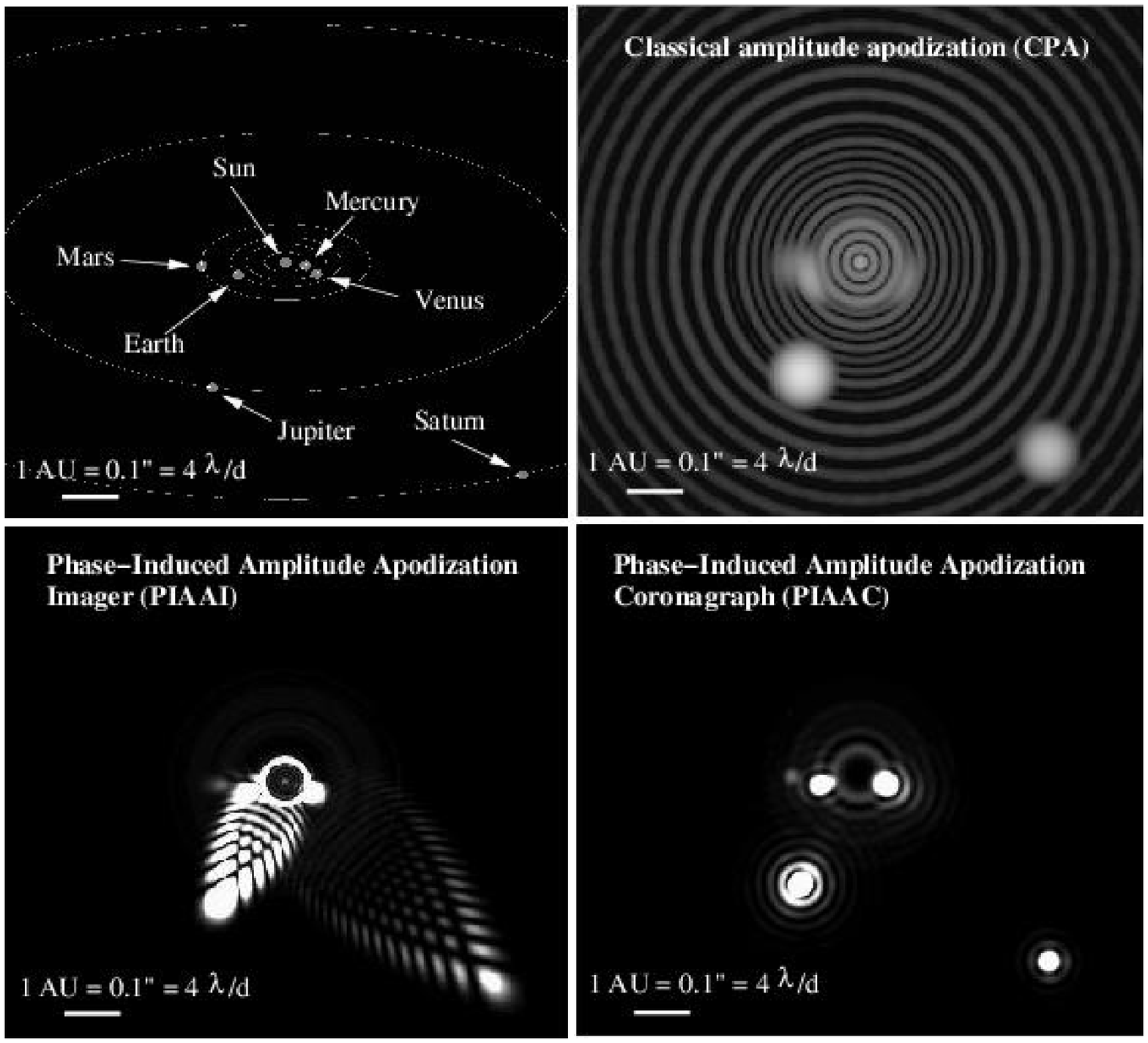}
\caption{\label{fig:solsyst} Simulated 0.5 $\mu$m images of the solar system at a distance of 10 pc with a 4m diameter telescope. The same apodization profile was produced by classical apodization (upper right) and PIAA (lower left). In the PIAAC (lower right) the original pupil geometry was restored before imaging. For the CPA and PIAA, an focal plane occulting mask was used to block the starlight, and the images shown are acquired in the reimaged focal plane. In all 3 images, the angular size of the star was taken into account and the faint concentric rings centered on the optical axis are residual starlight.}
\end{figure*}

Even in the absence of low order aberrations, the focal plane mask size needs to be increased because of the angular diameter of the central star. A Sun-like star at 5 parsec is 1 mas in radius. On a 4m telescope at $0.5 \mu m$, this corresponds to 0.04 $\lambda/d$. As shown in fig. \ref{fig:errsens}, with a pointing error equal to 0.04 $\lambda/d$, and a mask radius 40\% larger than the nominal size, the contrast level is better (by about a factor 10) than the $10^{10}$ requirement at all angular separations larger than the IWA. 

This 40\% increase in the mask size increases the IWA by 50\%, from 1.2 $\lambda/d$ to 1.8 $\lambda/d$. With the definitions of IWA and focal scale $(\lambda/d)_s$ adopted in this work, the radius of the occulting mask is approximately equal to the IWA of the PIAAC system.

The stars that have the largest angular diameter tend to be closer, and their habitable zone (the distance range from the central star where life as we know it might appear) is consequently at larger angular separation. There is therefore little penalty in increasing the IWA of the PIAAC by 50\% or so for those systems. However, the IWA should be kept small for distant (more than 10 pc) stars, and it would be advantageous to be able to ``tune'' the occulting mask size for each observation.

Since the PSF scale is wavelength-dependent, the mask size should be chosen to accommodate the longest wavelength seen by the detector. Since there is no penalty (other than an increase of IWA) in increasing the mask size, the contrast level is still maintained over the full spectral bandwidth, and the PIAAC does not suffer from chromatism as many coronagraphs do. 

\subsection{Sensitivity of the PIAAC for Earth-size planet detection}

As illustrated in Figure \ref{fig:solsyst} (lower right), the PIAAC delivers high quality images, with good angular resolution (equal to the diffraction limit of the telescope used) and very low distortion across the field of view necessary to image extrasolar planetary systems. In comparison, classical apodization (Figure \ref{fig:solsyst}, upper right) suffers from a large IWA, and a poor angular resolution, and the PIAAI (Figure \ref{fig:solsyst}, lower left) delivers aberrated images which are especially problematic in systems with multiple planets and/or complex exozodiacal light structures.

\begin{figure}[htb]
\includegraphics[scale=0.8]{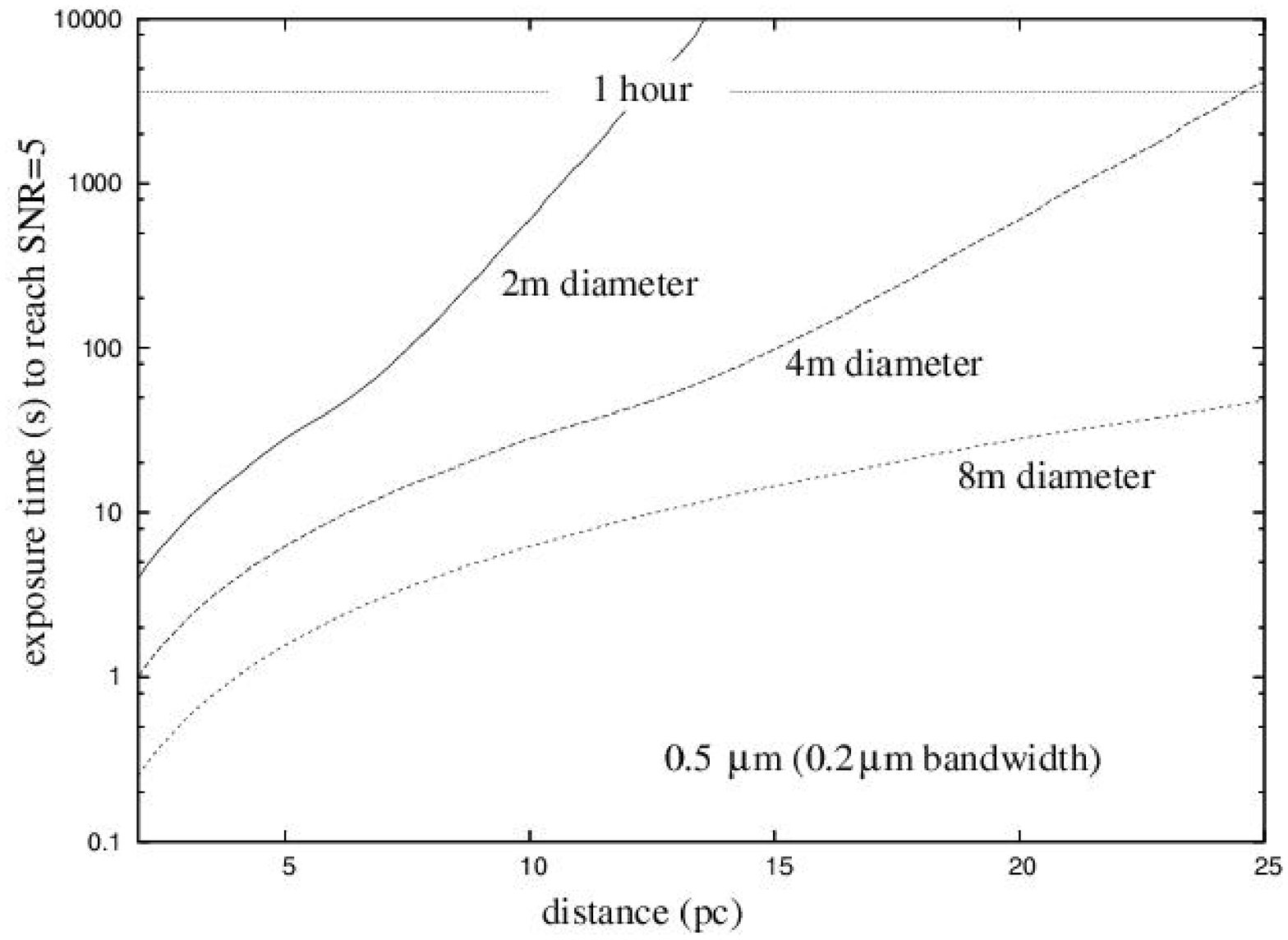}
\caption{\label{fig:ssetime1} Exposure time required to detect (SNR=5) an Earth-size planet at maximum elongation as a function of distance to the system, for 3 sizes of telescope (2, 4 and 8m diameter).}
\end{figure}

To test the sensitivity of the PIAAC for Earth-size planets detection, we consider a Sun-Earth system at maximum elongation. The projected separation is 1 AU, and the planet luminosity is $3.3\:10^{-10}$ times the central star luminosity. To compute the signal-to-noise ratio (SNR), we consider only photon noise from the planet and the starlight (noiseless detector). We adopt a focal plane mask size of $1.6 \lambda/d_s$, which provides a good level of immunity to pointing and focus errors (see \S\ref{ssec:ttsens}), and the stellar leak is estimated from the stellar angular diameter. The transmission of the coronagraph is also taken into account to estimate the planet flux in the focal plane detector. A central wavelength of 0.5$\mu$m and a spectral bandwidth of 0.2$\mu$m are adopted.

With this simple model, the exposure time required to reach a given SNR can be expressed as a function of the telescope diameter and the distance to the planetary system. As shown in Figure \ref{fig:ssetime1}, a 2m telescope can detect an Earth at 10 pc in 10 min, while it requires 28s on a 4m telescope and 6s on an 8m telescope. With a maximum allowed exposure time of 1 hour per target, Earth-size planets can be detected to 12.3 pc with a 2m telescope, and to 24.6 pc with a 4m telescope. Detection times are very short with an 8m telescope, even at 25 pc (53s). This simple model does not account for planet phase, likelihood of the planet to be near the maximum elongation, and stars of a different spectral type than the Sun. However, the results obtained clearly show that the PIAAC is a powerful tool for the direct detection of exoplanets, even with a 2m to 4m diameter telescope.

A thorough simulation of the PIAAC performance in carrying out a survey of nearby solar-type stars is in preparation and will be published separately under the same title : Part II. Performance.

\begin{figure}[htb]
\includegraphics[scale=0.8]{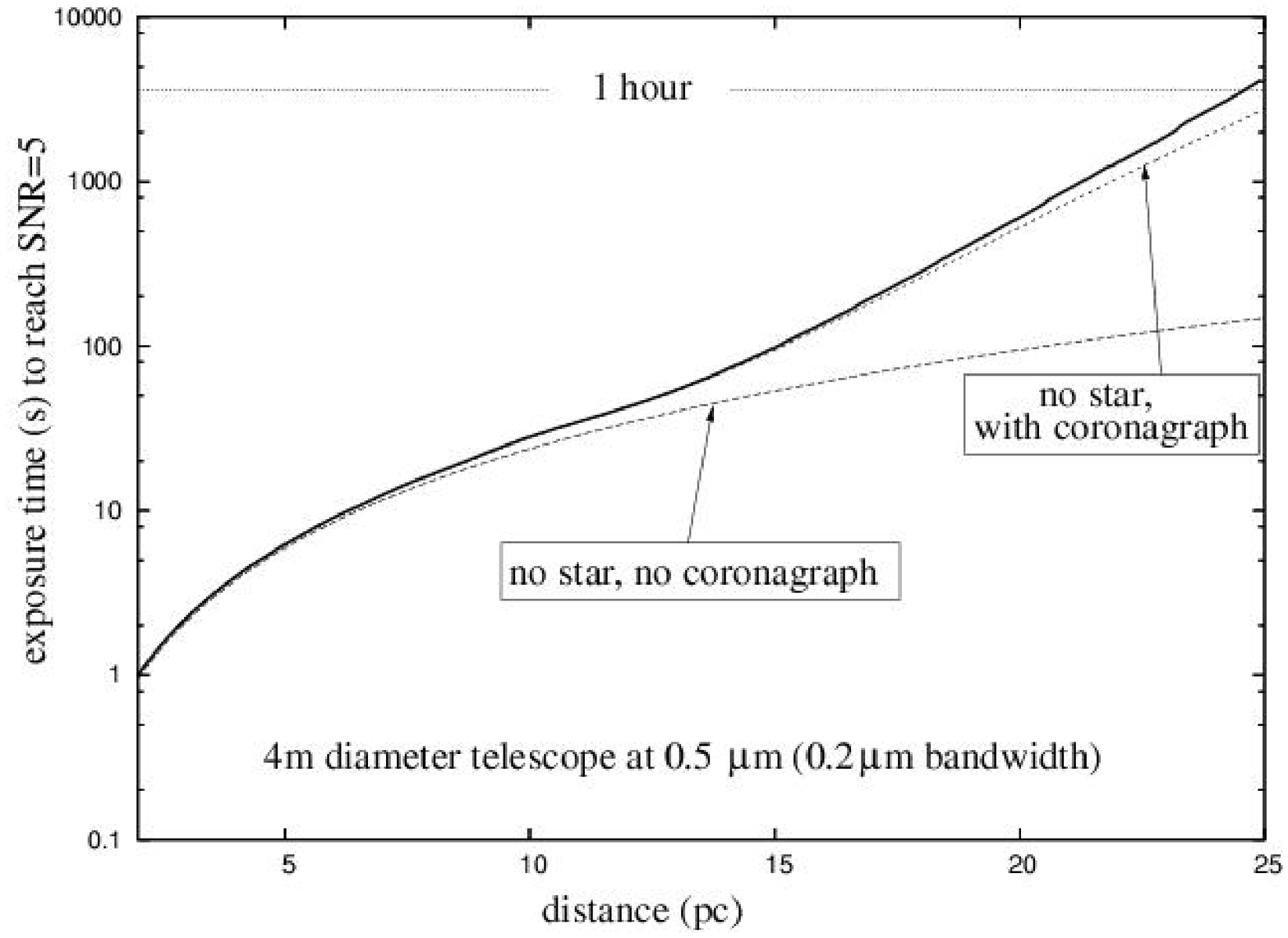}
\caption{\label{fig:ssetime} Exposure time required to detect (SNR=5) an Earth-size planet at maximum elongation as a function of distance to the system (full line). The same exposure time is also shown as if the residual starlight were perfectly removed (upper dashed line), and as if both the star and the coronagraph were removed (lower dashed line).}
\end{figure}

With a 4m telescope, the performance of the PIAAC is limited by the photon noise of the planet itself: the amount of residual starlight is smaller than the planet flux. This is illustrated in Figure \ref{fig:ssetime}, where the actual exposure time curve follows very closely the one computed with no central star. The coronagraph throughput is close to 100\% for systems closer than 15pc, and the sensitivity is therefore very close to what it would be if the planet were observed without its parent star and without a coronagraph (``no star, no coronagraph'' curve in Figure \ref{fig:ssetime}). At distances exceeding 15pc, the coronagraph throughput affects the planet flux, and therefore increases the exposure time required for detection. In this example, the IWA of the PIAAC corresponds to 1 AU at 25 pc, a separation for which the exposure time is 10 times longer due to the low throughput.

Although exozodiacal light (or other sources of ``background'' light) was not considered in this model, its flux would likely exceed the planet's flux inside the diffraction peak of the planet's image. In an exo-zodi limited measurement, the exposure time required for detection is proportional to the inverse of the exo-zodi flux contained inside the diffraction peak. The diffraction limit of the PIAAC is about 3 better than for a classical apodizer, which represents a gain of a factor about 10 in exposure time. Since classical apodizers transmit about 10\% to 20\% of the light gathered by the telescope, it is expected that the PIAAC requires exposure times 50 to 100 shorter than classical apodizers in the exo-zodi limited case.

\subsection{Use in space and on ground-based telescopes}
In this study, we have so far only considered the use of the PIAAC in a telescope which delivers a perfect wavefront. Mid and high spatial frequency wavefront aberrations will degrade the achievable image contrast in the PIAAC just like in other coronagraphs. In a well engineered space telescope, the drifts in the optics shapes can be small and slow, and wavefront control techniques could be used to bring the phase errors to an acceptable level (less than 1nm).

On ground based telescopes equipped with high-order AO systems, the wavefront aberrations cannot be reduced to such levels because of the speed of the turbulence and the limited number of photons available to measure it. Even with a ``perfect'' coronagraph, the achievable contrast level on these telescopes is therefore not very high (at best $10^7$ at a few $\lambda/d$ for very bright stars). The PIAAC however remains a very attractive coronagraph for ground based telescopes: it is achromatic, has a small IWA and a high throughput. Of particular interest is the ability to work at very small separations with a low sensitivity to tip-tilt errors in the wavefront (tip-tilt residuals are the dominant source of ``leak'' in coronagraphs with small IWAs). We note that plans to detect exo-planets with ground-based telescopes are based on very large apertures, working at a large multiple of $\lambda/d$. In these conditions, the telescopes will be typically limited not by the static diffraction in the PSF, but by the residual wavefront error. At large working angle and (relatively) large wavefront error, the PIAAC PSF benefits may not be important. The other PIAAC benefits of throughput and resolution will still be useful at relatively large working angle. The static PSF will be important for programs which require the small inner working angle.

\section{Conclusion}

The PIAAC offers a significant improvement over the previously proposed PIAAI concept \cite{guyo03}, as illustrated in fig. \ref{fig:solsyst}. The PIAAC combines high contrast, small IWA, full throughput and low sensitivity to pointing errors. Since the apodization is achieved by geometrical optics, it is achromatic and not likely to be prone to unwanted diffraction effects. No other current coronagraph design combines all these advantages. For example, classical coronagraphs that have small IWA (about $1 \lambda/d$) are very sensitive to pointing errors, and apodization-based coronagraphs are not sensitive to pointing but have large IWA and poor throughput. The PIAAC is therefore very well suited to image extrasolar planets, and can efficiently sample a large number of nearby stars with a moderate-size telescope. The performance (IWA, sensitivity and size of off-axis PSFs) of a PIAAC is equivalent to the performance of a classical pupil apodized telescope 2 to 3 times larger in diameter (see fig. \ref{fig:solsyst}).

\acknowledgements
This work was carried out under JPL contract numbers 1254445 and 1257767 for Development of Technologies for the Terrestrial Planet Finder Mission, with the support and hospitality of the National Astronomical Observatory of Japan.

\end{document}